# Rough Fabry-Perot cavity: a vastly multi-scale numerical problem

Tetiana Slipchenko,[1#] Jaime Abad-Arredondo,[2†#] Antonio Consoli,[1,3] Francisco J García Vidal,[2] Antonio I Fernández-Domínguez,[2] Pedro David García,[1] Cefe López[1,§]

[1]Instituto de Ciencia de Materiales de Madrid (ICMM), Consejo Superior de Investigaciones Científicas (CSIC), Calle Sor Juana Inés de la Cruz, 3, 28049 Madrid, Spain.
[2]Dep. Física Teórica de la materia Condensada, Universidad Autónoma de Madrid, Tomás y Valiente, 28049 Madrid and IFIMAC
[3]Escuela de Ingeniería de Fuenlabrada (EIF), Universidad Rey Juan Carlos (URJC), Camino del Molino 5, 28942 Fuenlabrada, Madrid, Spain
† *jaime.abad@uam.es*; § *c.lopez@csic.es*
# Both authors contributed equally to this work.

A commercial Fabry-Perot laser diode is characterized by highly disproportionate dimensions, which poses a significant numerical challenge, even for state-of-the-art tools. This challenge is exacerbated when one of the cavity mirrors is roughened, as is the case when fabricating random laser diodes. Such a system involves length scales from several hundred micrometres (length) to a few nanometres (roughness) all of which are relevant when studying optical properties in the visible. While involving an extreme range of dimensions, these cavities cannot be treated through statistical approaches such as those used with self-similar fractal structures known to show well-studied properties. Here we deploy numerical methods to compute cavity modes and show how random corrugations of the Fabry-Perot cavity wall affect statistical properties of their spectral features. Our study constitutes a necessary first step in developing technologically essential devices for photonic computation and efficient speckle-free illumination.

## 1. INTRODUCTION

Diffusive elements in optics can be seen as detrimental features as they typically lead to increased attenuation, uncontrolled light propagation, and reduced coherence [1,2]. However, it can be advantageous for the realization of non-conventional light emitters, like random lasers (RLs) [3], in which scattering elements provide optical feedback for lasing.

Random lasers have been firstly proposed as "non-resonant" [4] lasers in which one of the mirrors of a Fabry-Perot cavity (FPC) was substituted for a scattering surface. First observations of lasing emission have been obtained in dye-doped colloidal suspensions [5] and in semiconductor powders, [6] *i.e.*, in structures where the feedback is spatially distributed via a random spatial distribution of the refractive index.

Distributed feedback being the most common architecture considered in experimental realizations, theoretical investigations have been oriented to this kind of devices. Several theories [7] have been proposed approaching the problem as an exceptional case that can be understood with assumptions specific to RLs, *e. g.*, open cavities with quasi-bound states [8], extended amplified modes [9], closed feedback loops [10] or Anderson localization [11].

We recently proposed a unified taxonomy of lasing cavities, to include RLs, based on the nature of the cavity and the form of feedback: ordered/disordered structures and distributed/non-distributed feedback. This allows to classify them in four categories: FPC lasers, distributed feedback lasers and RLs with distributed and non-distributed feedback (see Supplementary Material in ref. [12]). We inferred this vision from the observation of the typical spectral emission of RLs from structures consisting of the active material, enclosed between two scattering elements [13] or a scattering element and a mirror [14]. Other authors also reported RLs with spatially separated regions for gain and feedback, *e.g.*, in active fibres with rough cladding [15], dried dye droplets with cracks [16] and polymers with scattering surfaces [17]. Previously referred to as RLs with spatially localized feedback, we propose the nomenclature of non-distributed feedback RLs or more generally of diffusive cavity lasers.

This class of emitters shows specific emission properties, which arise from the high numbers of optical modes existing in the cavity: multiple randomly distributed frequency peaks or single peak emission (interpreted as highly coupled modes), low spatial coherence and high divergence angle of emission. Different applications take advantage of these properties, *e.g.* high efficiency lighting [18], super spectral resolution spectroscopy [19], signal processing [20] and coding [21], photonic hardware for artificial intelligence [22], counterfeiting [23], and biomarkers [24]. However, a clear and solid theoretical treatment of this class of lasers is still missing.

Here, we offer a theoretical analysis and perform numerical simulations of the structure consisting of a plain mirror and a scattering surface, like the one based on semiconductor materials obtained in previous work [12].

## 2. PROBLEM DESCRIPTION

Semiconductor laser diodes rely on a carefully engineered structure designed to achieve efficient operation. The active medium, typically a multiple quantum well, is sandwiched between two doped layers that enable carrier injection. These layers, having a wider electronic band-gap and lower refractive index, also serve as confining regions for both charge carriers and optical modes. The device is terminated by cleaved edges, forming a FPC that provides strong feedback for in-plane propagation.

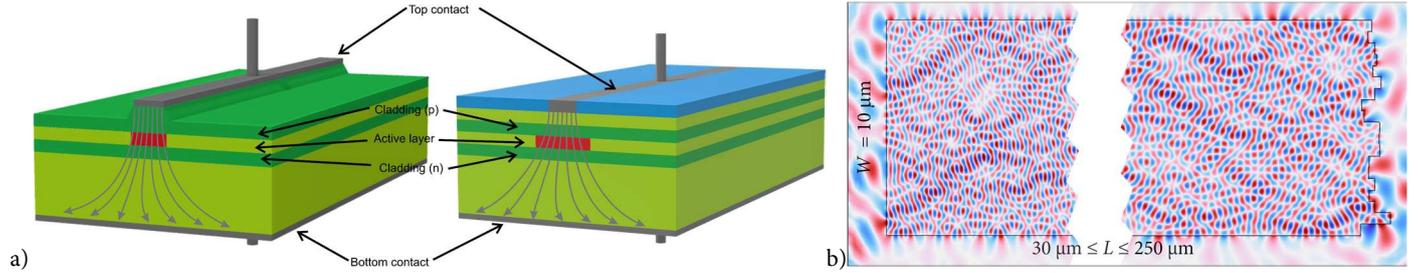

**Fig. 1** a) General design of a semiconductor laser diode. Depending on the width and type of lateral confinement several morphologies can be conceived. b) Geometry of the FP cavity and simulation domain including detail of *the* rough FP edge composed of $N_{sec}$ sections, each being assigned a random length uniformly fluctuating around a mean value.

Two common FPC configurations help define lasing properties. In electrode-controlled devices, a top electrode selectively activates the lasing region, creating an area where current density exceeds the necessary threshold (see **Fig. 1a** right drawing). An alternative approach confines the active medium in a ridge waveguide, preventing lateral current spread while enhancing both optical and electronic confinement (see Fig. a left drawing). The system considered in this study consists of a high aspect ratio 3D FPC, with active layers only a few nanometres thick compared to device width of tens or length of several hundred micrometres. This dimensional imbalance allows to model the structure effectively as two-dimensional.

Even under the simplifying assumption of a two-dimensional geometry, the large in-plane dimensions of commercial-scale FPCs (on the order of tens of thousands squared wavelengths) make numerical treatment highly challenging. Beyond pure scale considerations, a further challenge in laser cavity modelling is the impact of random surface roughness, since accurately describing the surface corrugations significantly increases the degrees of freedom of the problem.

To provide context, finite-element method eigensolvers typically depend on sparse-matrix diagonalization algorithms, such as the Arnoldi method implemented in ARPACK [25]. When applied to large-scale optical cavities with few relevant modes, these algorithms exhibit quadratic scaling in both time and memory with respect to the number of degrees of freedom [26,27]. Thus, it becomes apparent how precisely modelling laser cavities with lengths of tens or hundreds of micrometres, with corrugations in the end mirrors featuring correlation lengths of hundreds of nanometres and depths of tens of nanometres, requires handling very different length scales with high accuracy. In this work, we employ the finite-element method solvers implemented in COMSOL® Multiphysics to address these numerical challenges and compute the eigenmodes of the system.

We emphasize that in this study we focus on the passive (i.e., without gain) cavity modes of rough FPCs, which can offer valuable insight into behavior in the lasing regime. In these disordered cavities, optical modes initially experience gain compensation of roughness-induced scattering, followed by gain narrowing as they evolve toward the lasing state [28]. These lasing modes are defined through nonlinear interactions with the gain medium, and their direct numerical study remains a highly demanding task. Nevertheless, although lasing states do not strictly coincide with passive cavity solutions, they can be understood as emerging from them. Thus, analyzing the passive eigenmode spectrum offers a reasonable estimate of lasing frequency distributions. Furthermore, since these cavities exhibit inherent randomness, a statistical framework that describes their spectral properties and disorder-induced effects will also be valuable to describe the properties of lasing modes.

### 3. NUMERICAL INSPECTION

To study the resonant modes and frequency characteristics of passive FPCs, with a rough mirror, we perform an eigenfrequency analysis using COMSOL Multiphysics. The FPCs considered here (shown in Fig. 1a) are dielectric slabs with refractive index $n_s$ = 3.86 embedded in air. The slab has a fixed width, $W$ = 10 μm, and we studied cases with a length, $L$, ranging from 30 μm to 250 μm. To introduce surface roughness in a controlled way while holding computational cost in check, we applied a random corrugation to the right-hand side boundary of the dielectric slab. This corrugation is in the form of a simple crenelated profile consisting of $N_{sec}$ sections of different length (See **Fig. 1b**). All sections have the same width but vary in length leading to steps of varying depth in the edge. These depths are drawn from a uniform random distribution in the interval [0,$h$], where $h$ defines the maximum fluctuation amplitude considered in a realisation. For the eigenfrequency analysis, we specified an eigenvalue search range centred around 500 THz and solved for the 50 closest eigenfrequencies, allowing us to capture a comprehensive set of resonant modes near this frequency. This setup provides insights into the cavity's mode structure and frequency response under different levels of boundary roughness. All results discussed in this work pertain to transverse electric (TE) polarization. Similar results are obtained for transverse magnetic (TM) polarization as shown in the Supplementary Material.

#### A. Average spectral distance

The average spectral separation, $\langle s \rangle$, a dimensionless spacing between adjacent eigenfrequencies, is calculated as a normalised average of the mode spacings $\Delta f_n = f_{n+1} - f_n$:

$$\langle s \rangle = \frac{1}{N} \sum_{n=1}^{N-1} \frac{f_{n+1} - f_n}{f_n} \quad (1)$$

where $N$ is the number of eigenfrequencies considered, $f_n$, calculated in a given frequency interval.

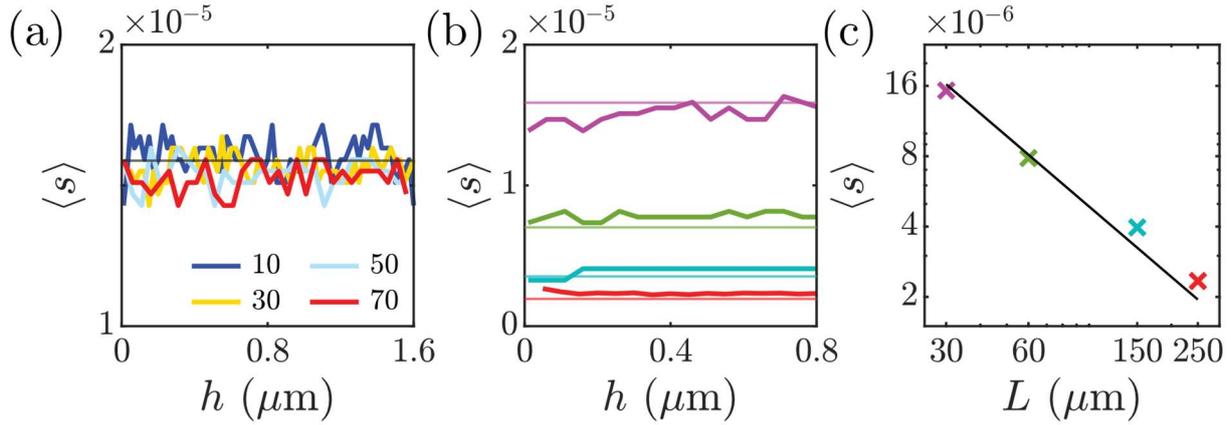

**Fig. 2.** a). Dimensionless average spectral separation, ⟨s⟩, as a function roughness depth, h, for different number of the sections $N_{sect}$ for a FPC with a randomly corrugated wall. b) Same as a function of roughness depth h for FPCs with various lengths L. Different coloured lines correspond to cavities with L = 30 (purple), 60 (green), 150 (blue) and 250 (red) μm, respectively. Horizontal lines correspond to ⟨s⟩ in the pristine case for the different cavity lengths. c) Mean (in h) average spectral distance as a function of FPC length in log-log scale. The black solid line represents the fit of the numerical data with analytical expression ⟨s⟩ = 4.87×10⁻¹⁰ / L. $N_{sect}$ = 40 for panels b and c.

**Fig. 2a** illustrates the average spectral separation, ⟨s⟩, as a function of roughness depth, h, of the corrugation and for different number of the sections, $N_{sect}$, for a cavity of length L = 30 μm. Across the entire range of $N_{sect}$ and h, ⟨s⟩ fluctuates around the baseline given by the free spectral range (FSR) of the regular FPC shown as a horizontal solid line. This indicates that disorder introduces variability in spectral spacing but generally does not lead to substantial modifications of the average spectral separation of the regular FPC. In **Fig. 2b**, we set $N_{sect}$ = 40 and show ⟨s⟩ as a function of h for cavities of lengths 30, 60, 150 and 250 μm, shown in purple, green, blue, and red respectively. The normalized averaged spectral separations again fluctuate around the FSR of the corresponding regular FPCs. These reference FSR values are indicated by horizontal lines. Since h does not seem to have a strong impact, we take the average for all h and plot ⟨s⟩ as a function of L in **Fig. 2c**. One can see that ⟨s⟩ decreases as L increases. This reflects the expected inverse proportionality between the FSR of a FPC and its length, $f_{FSR} \propto 1/L$. Under that assumption, we fit the numerical data for mean (in h) average spectral separation to an analytical expression, finding ⟨s⟩ = 4.87×10⁻¹⁰ / L. The analytical results correspond to the black solid line. They show good agreement with the numerical data, represented by crosses, over the considered range.

### B. Average spectral broadening:

The dimensionless mode spectral broadening—indicating the spread of resonance linewidths— can be calculated as the average of the modes' inverse quality factors:

$$\langle Q^{-1} \rangle = \frac{1}{N} \sum_{n=1}^{N} \frac{1}{Q_n} \quad (2)$$

where $Q_n = f_n/\delta f_n$ with $\delta f_n$ the mode width. In **Fig. 3a** we show that ⟨$Q^{-1}$⟩ initially sharply increases with h for any number of sections (cavity of length L = 30 μm). Beyond a certain threshold, around h ≈ 0.5 μm, ⟨$Q^{-1}$⟩ values begin to stabilize, particularly for the larger numbers of sections. This plateau suggests that, after an initial high sensitivity to roughness, the spectral broadening reaches a level where further increases in h does not lead to a proportional increase in ⟨$Q^{-1}$⟩. In **Fig. 3b**, the number of sections is fixed to $N_{sect}$ = 40, and we show ⟨$Q^{-1}$⟩ as a function of h for cavity lengths of 30, 60, 150 and 250 μm, given respectively by the purple, green, blue, and red lines. All traces follow the same structure, with a sharp initial rise and then a plateau onto a stable value, indicating that broadening saturates for all cavity sizes. In **Fig. 3c** we plot the estimated asymptotic broadening value for the different cavities in Fig. 3b. This figure illustrates that ⟨$Q^{-1}$⟩ decreases as L increases, indicating that quality factor reduction is related to the quality factor of the pristine cavity. Fig. 3c further reinforces the inverse relationship between L and ⟨$Q^{-1}$⟩ displaying a clear downward trend as L increases. This inverse mean length dependence suggests that longer cavities are more resilient to mode broadening effects, probably due to more efficient confinement and reduced sensitivity to structural disorder. In principle, in longer cavities with flat mirrors, quality factors are larger, and therefore the modes, better confined, are less sensitive to mirror corrugation.

### 4. ANALYTICAL MODEL

In the previous section, we demonstrated that roughness leads to a reduction of the quality factor of resonances that saturates to a limiting non-zero value, while retaining the free spectral range of the unmodified cavities, *i.e.* without the introduction or removal of modes. In order to shed light onto this behaviour, we now introduce a simple analytical approach based on a 1D cavity, with one pristine mirror, with reflection coefficient r, and one rough mirror. The core assumption behind the model is that upon incidence on the rough mirror, waves will be reflected in a superposition of different phase-offsets, ϕ, where the weight of the different phase components is *normally* distributed around zero phase offset with $\sigma_\phi$ standard deviation in a way similar to ref. [13]. Here we are assuming that the magnitude of the amplitude coefficient for reflection, r, is merely determined by the index contrast and the phase change upon reflection, ϕ, obeys a Gaussian distribution of variance $\sigma_\phi$. That is to say, the probability with which an incident wavefront is reflected with a phase ϕ is given by $P(\phi) = \frac{1}{\sqrt{2\pi\sigma^2}} e^{-\phi^2/2\sigma_\phi^2}$. Under this assumption computing the expected value of $e^{-i\phi}$ is equivalent to calculating the Fourier transform of a Gaussian which is a Gaussian. So, it can be shown (see Supplementary Material) that the expected value for the amplitude

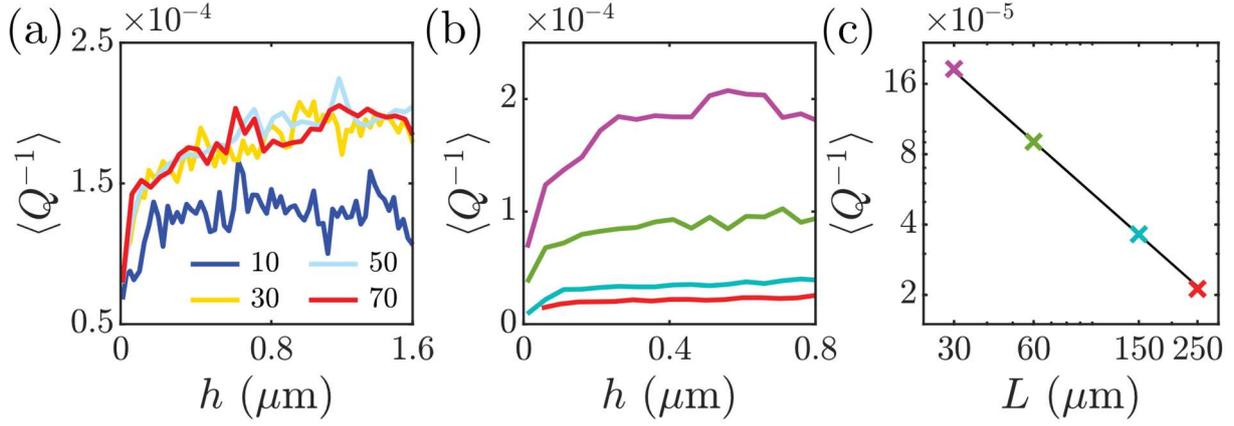

**Fig. 3** a). Dimensionless average spectral broadening ⟨$Q^{-1}$⟩ as a function of roughness depth, $h$, for different number of the sections $N_{sect}$. Cavity length is 30 μm. b) Dimensionless average spectral broadening as a function of roughness depth for different FPC lengths, $L$: 30 μm (purple), 60 μm (green), 150 μm (blue) and 250 μm (red) respectively. c) ⟨$Q^{-1}$⟩ averaged over roughness depth ($h$), as a function of FPC length, shown on a log-log scale. $N_{sect}$ = 40 for panels b) and c). The black solid line represents the fit of the numerical data analytical expression ⟨$Q^{-1}$⟩ = 5.43×10$^{-3}$/$L$.

reflectance coefficient is $r' = re^{-\frac{1}{2}\sigma_\phi^2}$, which shows that this phase supperposition, through destructive interference of the reflected fields, effectively acts as a loss mechanism for the FPC. We further assume that the standard deviation of the phase distribution can be related to the surface's RMS roughness, $\sigma_h$, by, $\sigma_\phi = \alpha k \sigma_h$ where $k = 2\pi n_s/\lambda$ is the wavevector in the cavity and $\alpha$ is an adimensional factor related to the sensitivity of the particular mode to the roughness.

To test this model, we perform numerical simulations in COMSOL measuring the power reflected from a rough surface. In sight of Fig. 2c and Fig. 3c we believe numerical estimates can be safely extrapolated to large $L$ so, throughout this section, in order to study the statistical properties of these cavities, we will focus on FPCs of smaller size. This alleviates the massive computational effort required to model roughness in a more realistic manner, closer to experimental realizations. In particular, roughness is implemented by superposition of random waves with correlation lengths of 100 nm and a controlled root mean square (RMS) displacement. **Fig. 4** presents a statistical analysis of the reflected power from a rough surface, based on 1000 realizations per roughness level. In the panels, we distinguish between specular ($R_0$), and total integrated reflectance ($R_T$), obtained for a plane wave of $\lambda = 600$ nm. The wave is normally incident from a medium with $n_s = 3.86$ onto the rough surface at the boundary with vacuum. The colourmaps render histograms of the numerically-obtained reflectivity values, ranging from white (least frequency, minimum number of counts) to black (highest frequency, maximum number of counts) for both quantities. These panels enable us to infer the probability distribution for the underlying stochastic variable and determine the most likely values for $R_0$ and $R_T$ as a function of the mirror roughness. In **Fig. 4a**, and in the limit of low roughness, the reflectance is well described by the Fresnel reflection coefficient (horizontal dashed line). However, as roughness is increased beyond $10^{-2}\lambda$, specular reflectance values drop off significantly, reaching vanishing values for roughness levels of $10^{-1}\lambda$. In **Fig. 4b** we plot the specular reflectance in logarithmic scale to reveal that it plateaus to a limiting value at high roughness levels. To capture this behaviour, we include a phenomenological remnant reflectance, $r_0$, in our model, which will dominate the rough-cavity limit. Therefore, we write the reflectance from the rough surface as

$$r' = \left|(r - r_0)e^{-\frac{1}{2}(\alpha k \sigma_h)^2} + r_0\right|^2 \quad (3)$$

In Fig. 4 we show the fitting (using 2 fitting parameters, $r_0$ and $\alpha$) of the simulated reflected values to this distribution as a green line in both panels, demonstrating that this phenomenological approach correctly captures the degradation of the feedback mechanism that will ultimately lead to the general broadening of cavity resonances showcased in the previous section. The best fit in this case corresponds to $r_0 = 0.16$ and $\alpha = 1.87$, which predicts a residual reflectance of the rough mirror of $|r_0|^2 = 2\cdot 10^{-2}$. Furthermore, note that for the simple case of normal incidence, the difference of optical path for a wave impinging on a boundary offset by $\sigma_h$ would be of approximately $2\sigma_h$ and, therefore, from simple geometric arguments, one could expect a value of the sensitivity parameter, $\alpha$ close to 2, as we find in the fit.

Once the reflection coefficient from the single scattering event is stablished, as shown in the Supplementary Material, we use it to develop an analytical solution for the fields scattered by a 1D cavity in which one mirror is pristine, with coefficient $r$, and the other mirror is roughened, with reflection coefficient $r'$. The treatment can be readily extended to more realistic Fabry-Perot cavities, where the front and back mirrors may have differing reflectivities, as is commonly the case in conventional diode lasers. Based on this analytical model, we can determine the resonant modes of the cavity, and by performing a small frequency expansion around these resonances, one may find the usual associated Lorentzian lineshapes, and also an expression for the quality factor of such resonances. See Supplemental Material for details. We can write the expression for the quality factor of the $n$-th cavity mode in the following compact form:

$$Q_n = \frac{Q_n^{(P)}}{\sqrt{1 + \frac{(1-\delta)}{2\delta}[F_n^{(+)} - F_n^{(-)}\delta]}} \quad (4)$$

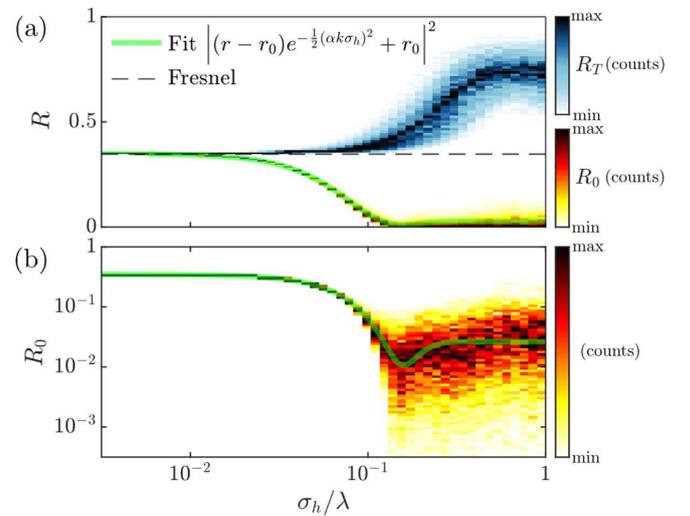

**Fig. 4.** Statistical study of reflectance from a rough flat surface. a) 2d histograms of total (blue) and specular (red) reflectance from a rough surface. (1k realizations per roughness level) together with a fitting to analytical expression. For each roughness level, the most frequent value is shown in black and the least in white for both magnitudes. b) Histogram of specular reflectance now in log-scale, revealing the saturation of the value of the specular reflectivity for large roughness.

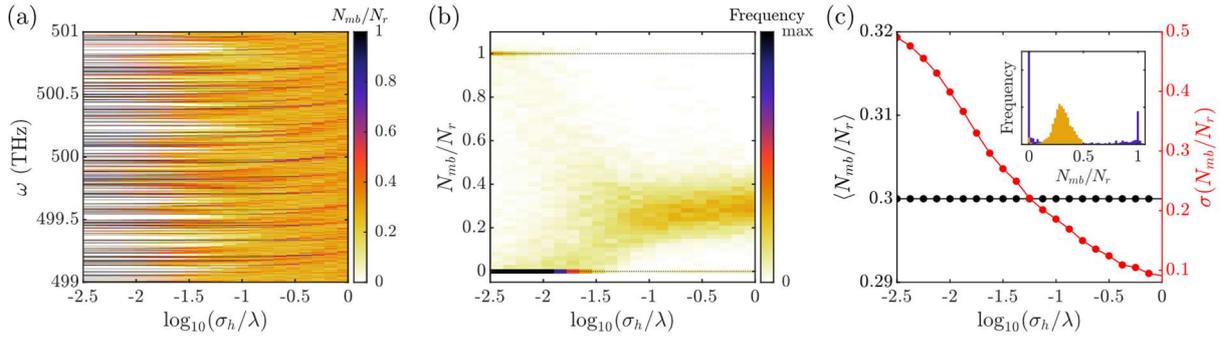

**Fig. 5.** Statistical analysis of the spectral properties of rough FP cavities. Panels a-c correspond to cavities of size 10 × 30 micrometers. a) Two-dimensional map presenting the number of modes in the whole realisations set per frequency interval, normalized to number of cavity realizations as a function of frequency and roughness. b) Frequency or number of occurrences in which the bins of panel a contain $N_{mb}/N_r$ modes as a function of roughness. Colour scale saturates to better showcase a feature at $N_{mb}/N_r \approx 0.3$. c) Ensemble-averaged and normalised number of modes per frequency interval (bin), and standard deviation. The inset shows line cuts of the distribution of panel b for roughness values $\log(\sigma_h/\lambda)$ = 2.5 (purple) and 0 (yellow).

In this expression, $Q_n^{(P)}$ is the quality factor in the pristine limit and $F_\pm$ are defined as $F_n^{(\pm)} \equiv (1 + A_n^2 \pm A_n\sqrt{2 + A_n^2})/A_n^2$, where $A_n = (\sqrt{2}n\pi)/Q_n^{(P)}$ is related to the mode's order, $n$, and unperturbed quality factor. Finally, the variable in Eq. (4) that introduces the effect of roughness is the $\delta$ factor, defined as

$$\delta = e^{-\frac{\sigma_\phi^2}{4}}\left[\cosh\left(\frac{\sigma_\phi^2}{4}\right) - \sinh\left(\frac{\sigma_\phi^2}{4}\right)(1 - 2\beta)\right] \quad (5)$$

which depends on the roughness level through $\sigma_\phi = \alpha k \sigma_h$, as discussed above. Furthermore, note that this factor also contains the effect of saturation of reflectance losses through $\beta \equiv r_0/r$. This term leads to a saturating behaviour in the quality factor too, as observed in Fig. 4 which, in the limit of high roughness, tends to a saturation value obtained from Eq. (4) by replacing $\delta$ by $\beta$. This directly reproduces the numerically observed fact that the saturated $Q$ factor for the rough cavities is proportional to cavity length (since the quality factor in the pristine case follows this proportionality).

Note that in this model, although the quality factor is modified, as compared to the pristine case, the resonance frequency of the different modes is not affected. This is a direct consequence of the 1D, single mode nature of the model, and constitutes a significant simplification. More realistic characterization would involve considering how roughness mixes different modes of the 3D cavity, which would lead to mode hybridization and frequency shifts [29]. Nevertheless, as we will show in the next section, the exact resonance frequency of a given mode is not particularly relevant for the statistical analysis. As shown in the Supplemental Material, a first-order perturbative treatment predicts that the expectation values of the resonance frequencies remain unchanged to leading order in roughness. For this reason, our simplified model adequately captures the degradation of resonances resulting from complex mode mixing.

## 5. STATISTICAL PROPERTIES OF ROUGH FPC

In what follows, we present a statistical study of 2D FPCs of fixed size, 10 μm by 30 μm, in which roughness is modelled through a random corrugation of the same characteristics as used for Fig 4. To obtain the data we generate $N_r$ = 50 random realizations of cavities for each roughness level and record the eigenfrequencies in a 2 THz interval around a central frequency of 500 THz. We will follow the same structure as Section 2: first looking at the properties of the spectral distribution and then proceeding to analyse the effect that roughness has on the quality factors of the modes.

In **Fig. 5a** we show a 2D histogram where we divide the 2 THz frequency window into one-thousand bins of 2 GHz width (which is narrower than the pristine cavity's FSR of approximately 7 GHz implying that roughly three out of four bins must be empty in a typical realization). For each frequency bin and each roughness level, we plot the number of modes in the bin, $N_{mb}$, as a function of roughness averaged over realizations. Thus $N_{mb}/N_r$ allows us to infer the probability that FPC contains a mode within a 7 GHz window around the central frequency of the bin. Note that here we are using that there is no FPCs presenting more than 1 mode within a single frequency window. For low roughness levels, the mode distribution is sharply structured, always following the pristine case. As roughness is increased, these eigenfrequencies undergo slight shifts in spectral position and broadening, leading to a smearing out of the initially spiky lineshape. Further increasing the roughness leads to ever smoother spectra, which tends to a uniform distribution in the high roughness limit. The modal spectrum variation with increasing roughness can be seen in the colour map in **Fig. 5b**. It shows the frequency with which bins contain a certain number of modes $N_{mb}/N_r$, as a function of the mirror roughness normalized to the number of cavity realizations. For very low levels of roughness, the underlying probability distribution is bimodal, with a dominant peak at $N_{mb}/N_r = 0$ and a small one at $N_{mb}/N_r = 1$. That is, in this regime, most of the frequency intervals are empty in all realisations; only a marginally few containing exactly 1 mode, while the vast majority lies in between.

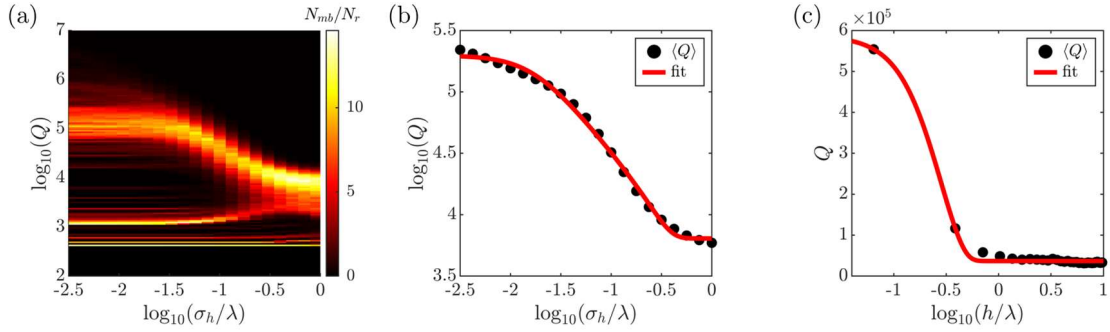

**Fig. 6.** Statistical analysis of rough FP cavities. a) Two-dimensional histogram showing the number of modes for a given roughness and quality factor. b) Mean quality factor as a function of roughness and fitting to the analytical model. Panels a-b correspond to cavities 10 × 30 microns. c) Same as panel b) for data corresponding to larger cavity of size 10 × 150 micrometers.

As roughness increases, more frequency windows in the whole realisation set contain modes, as shown by the smearing out of the initially sharp feature at $N_{mb}/N_r = 0$. Now, intermediate values of $N_{mb}/N_r$ appear with significant frequency, thereby reducing the incidence of zero or unity ratios. For high roughness, a pattern resembling a normal distribution emerges in the frequency map, with a mean value of $\langle N_{mb}/N_r \rangle \approx 0.3$ and a rather small standard deviation, $\sigma(N_{mb}/N_r)$, of the order of 0.1. Thus, in the rough-cavity limit, all frequency bins contain roughly the same number of modes, which agrees with a uniform frequency distribution at high $\sigma_h$ in Fig. 5a.

**Fig. 5c** presents the statistical analysis of the values of $N_{mb}/N_r$, across different roughness levels, rendering both its mean value and standard deviation. The mean value, $\langle N_{mb}/N_r \rangle$, remains constant despite variations in roughness, indicating that the total number of modes within the fixed frequency window (499-501 THz) is invariant, and thus showing that the FSR of the cavity is unaltered by roughness. In contrast, the standard deviation decreases as roughness increases. This should not be surprising considering that for the lowest roughness values the distribution is bimodal peaking a zero and one and maximizes the standard deviation $\sigma(N_{mb}/N_r)$.

At low roughness many frequency bins are empty because mode width is narrower than bin span, but as roughness grows the discrete mode spectra transitions into the rough cavity limit, where modes spread and frequency bins present a normally distributed occupancy. This trend highlights a transition in the behavior of the FPC spectral features, where increased disorder leads to statistical smoothing of spectral characteristics. Additionally, the inset of Fig. 5c presents line cuts of the $N_{mb}/N_r$ distribution extracted from Fig. 5b, illustrating the two limiting cases of $\sigma_h/\lambda = 10^{-2.5}$ and $\sigma_h/\lambda = 1$. These visualizations further emphasize the statistical evolution of mode distributions under increasing roughness.

We now shift our attention to the behaviour of the quality factor of the rough cavity modes. In **Fig. 6a** we show the histogram of the number of modes with a given quality factor for each roughness level. In this case, we divide the $Q$-factor axis into 150 bins, and display in colour the number of modes per bin, $N_{mb}$, normalized to the number of rough cavity realizations at each roughness level, $N_r = 50$. In this map one readily sees that, for low roughness levels, two sets of quality factor values appear frequently, which we associate to two groups of FPC modes: those with a very narrow distribution of low $Q$ values, and those with broader range of –two orders of magnitude— higher quality factors.

For both groups of modes, it is *apparent* that $Q$ is almost unmodified for $\sigma_h$ below ~$0.01\lambda$. Beyond that, the modes with highest quality factors, and therefore most sensitive to additional loss mechanisms, experience a clear drop in $Q$ (shortening of their lifetime), which ultimately saturates to a fixed value for $\sigma_h \approx \lambda$. Interestingly, the group of modes with $Q < 10^3$ is completely unfazed by the introduction of roughness. This can be understood by considering roughness as another loss mechanism: in the regime in which the mode broadening is governed by other phenomenon, *e.g.*, scattering or material absorption, the extra losses stemming from roughness may be negligible. Furthermore, one can also observe how some modes with initially low-$Q$ experience a slight increase in their quality factor. This we link to their roughness-induced hybridization with higher-$Q$ modes, with which they interact to form a low-$Q$ tail in the distribution at high roughness levels. As mentioned before, this cannot be modelled by our simple analytical treatment, since we are assuming fully independent modes. Nevertheless, this effect remains minor. In **Fig. 6b** we show the ensemble-averaged quality factor (consider all groups of modes introduced above), together with the fit to the analytical expression in Eq. (4), demonstrating an excellent agreement. Furthermore, in **Fig. 6c** we also fit the analytical expression to the average quality factors obtained from cavities larger than $L = 150$ μm (as those considered in Fig 2), also demonstrating a good agreement. These panels show that our simple analytical model can provide valuable insight into the modification of the FPC modes as roughness is introduced.

## 6. CONCLUSIONS

In summary, we have carried out numerical simulations of vastly multiscale FPCs involving dimensions from sub-millimetre (length) to nanometre (depth) scales to study the effect of random roughness with correlation lengths of the order of sub-micrometre. The simulations show that spectral mode distributions, allowing for the fluctuations inherent to successive realisations, follow approximately the perfect FPC free spectral ranges without much dependence on roughness depth. These also yield, as expected, a compression of the FSR approximately following a $1/L$ trend as the average length of the cavity grows. The quality factor of the modes, after a fast start, decrease slowly as the roughness of the output mirror increases.

Most of these features are captured by an analytical model that considers the reflection on the rough FPC mirror as a source of normally distributed phase that, through interference, permits treating the roughness as the origin of a reduced reflectance. Assuming that the phase distribution is associated to the roughness depth, this model allows to analytically compute effective reflectance and quality factors, all in very good agreement with numerical results. Roughness related loss of quality saturates leading to a residual value for very high roughness.

We expect that this kind of studies will allow a deeper understanding of the impact of fabrication on optical performance of devices [14] and guide future work.

### FUNDING

This work has received funding from MICIU Grants Ref. PID2021-124814NB-C21, PID2021-126964OB-I00, PID2021-127968NB-I00, and PDC2022-133418-I00. AC, PDG, and CL acknowledge support from the Spanish MICIU *Severo Ochoa* program for Centres of Excellence through grant CEX2024- 001445-S. JBA, FJGV and AIFD acknowledge funding from the *María de Maeztu* program for Units of Excellence in R&D through grant CEX2023-001316-M.

### DISCLOSURE

The authors declare no conflicts of interest.

The data that support the plots within this paper are available from the corresponding authors upon reasonable request. The Python codes developed to execute the calculations presented in this paper are available from the corresponding authors upon reasonable request.